\documentclass[adp,a4paper,fleqn%
]{w-art}
\usepackage{times,cite,w-thm}
\theoremstyle{plain}

\theoremstyle{definition}

\usepackage[]{graphicx}
\begin{document}
\DOIsuffix{theDOIsuffix}
\Volume{16}
\Month{01}
\Year{2007}
\pagespan{1}{}



\title[Collective Excitations of Impure Graphene]{Localised Magneto-Optical Collective Excitations of Impure Graphene}
\author[A.M.\ Fischer]{Andrea M.\ Fischer\inst{1,}%
  \footnote{Corresponding author\quad E-mail:~\textsf{A.M.Fischer@warwick.ac.uk},
            Phone: +44\,(24)\,765\,74752,
            Fax: +44\,(24)\,765\,73133}}
\address[\inst{1}]{Department of Physics and Centre for Scientific Computing,
  University of Warwick, Coventry CV4 7AL, United Kingdom}
\author[A.B.\ Dzyubenko]{Alexander B.\ Dzyubenko\inst{2,3}}
\address[\inst{2}]{Department of Physics, California State University Bakersfield, Bakersfield, CA 93311, USA}
\address[\inst{3}]{General Physics Institute, Russian Academy of Sciences, Moscow 119991, Russia}
\author[R.A.\ R\"omer]{Rudolf A.\ R\"omer\inst{1}}
\begin{abstract}
We study optically-induced collective excitations of graphene in the presence of a strong perpendicular magnetic field and a single impurity. We determine the energies and absorption strengths of these excitations, which become localised on the impurity. Two different types of impurity are considered i.\ the long-range Coulomb impurity, ii.\ a $\delta$-function impurity located at either an $A$ or $B$ graphene sublattice site. Both impurity types result in some bound states appearing both above and below the magnetoplasmon continuum, although the effect of the short-range impurity is less pronounced. The dependence of the energies and oscillator strengths of the bound states on the filling factor is investigated. 
\end{abstract}
\maketitle                   

\section{Introduction}
\label{intro}
Graphene, an atomically thick layer of graphite, has received a tremendous amount of attention since it was initially isolated five years ago \cite{CasGPNG09}. From a theoretical perspective, this is largely due to its linear dispersion relation at low energies, which occurs around two inequivalent corners of the Brillouin zone, the $\bf{K}$ and $\bf{K'}$ valleys. Close to these points, the single particle states have a spinor character and well defined chirality due to a pseudospin resulting from the inequivalent $A$ and $B$ sublattices of the honeycomb lattice. There has been much discussion in the literature regarding the role of electron-electron ($e$-$e$) interactions in graphene, particulary in relation to disorder \cite{CasGPNG09}. In undoped graphene the most striking effect of the long-range Coulomb $e$-$e$ interaction on the material's low energy properties, is simply a renormalisation of the Fermi velocity. This and the irrelevance of short-range $e$-$e$ interactions are well established results, which have promoted the idea that $e$-$e$ interactions are largely unimportant in graphene. However, recent work has portrayed it very differently, as a strongly interacting quantum liquid \cite{MulSF09}. In the present study, we examine the optically-induced excitations of monolayer graphene in a strong perpendicular magnetic field, which become localised on a single impurity with an axially-symmetric potential. Single particle excitations are mixed via the Coulomb interaction resulting in a collective excitation of the whole system. There is little discussion in the literature of the effect of a strong magnetic field upon the $e$-$e$ interactions, although optical properties have been suggested to be sensitive to them \cite{GruVV09}. They are treated here beyond the mean field level and are central to the problem, significantly altering even the qualitative nature of the bound states. We explore how these states are influenced by the filling factor and the nature of the impurity, considering a $\delta$-function scatterer and a Coulomb impurity. The effects of the light polarisation and the impurity strength are presented in Ref.\ \cite{FisDR09a,FisDR09b}. 
\section{Theoretical Model}
\label{sec-theory}
Let us first recall \cite{FisDR09a} the single particle problem in the absence of an impurity potential. A single electron wavefunction in, e.g.\ the ${\bf K}$ valley (pseudospin $\sigma=\Uparrow$),
is 
$\Psi_{n s \Uparrow m}(\mathbf{r})  = 
\langle \mathbf{r} |  c^{\dag}_{n s \Uparrow m} |0 \rangle 
 =\Phi_{n \Uparrow m}(\mathbf{r})\chi_s
                 = a_n
                     (\mathcal{S}_n \phi_{|n|- 1 \,  m}(\mathbf{r}),
                          \phi_{|n| \,  m}(\mathbf{r}),
                                     0,
                                     0)
\chi_s$,
where the symmetric gauge ${\bf A} = \frac12 {\bf B} \times {\bf r}$ is used to describe the perpendicular magnetic field.
Here, $n$ is an integer LL number, $\phi_{n  m}({\bf r})$ is a 2DEG wavefunction
with oscillator quantum number $m = 0, 1, \ldots$, $a_n=2^{\frac{1}{2}(\delta_{n,0} -1)}$,
$\mathcal{S}_n={\rm sign}(n)$ (with $\mathcal{S}_0=0$) and
$\chi_s$ is the spin part corresponding to two spin projections
$s = \uparrow, \downarrow$. The corresponding wavefunction in the ${\bf K'}$ valley ($\sigma=\Downarrow$) is obtained by reversing the order of the spinor components. We consider Landau levels (LLs), which are split into four non-degenerate sublevels, due to Zeeman splitting ($\hbar\omega_s$) and an additional valley pseudospin splitting ($\hbar\omega_{v}$) with $\hbar\omega_{v}\ll\hbar\omega_s$ (cf.\ Fig.\ \ref{fig-diag}). Valley splitting occurs in  high magnetic fields and has been seen clearly for the zeroth LL, with evidence for the $n=\pm1$ LLs observed, but somewhat weaker \cite{ZhaJSP06}. Defining the composite indices $\mathcal{N} = \{ n s \sigma \}$ and $N = \{ n \sigma m\}$, the single particle energy is $\epsilon_\mathcal{N} = \mathcal{S}_n \hbar\omega_c \sqrt{|n|} + \hbar\omega_s s_z + \hbar\omega_{v} \sigma_z$,
where $\hbar\omega_c =  v_F \sqrt{2 e \hbar B/c}$ is the cyclotron energy in graphene.

The impurity is described by an axially symmetric potential, which is given by $V_\mathrm{C}(\mathbf{r})= \pm e^2/\varepsilon_{\rm imp}|\mathbf{r}|$ for the Coulomb impurity, where $\varepsilon_{\rm imp}$ is the effective dielectric constant for the electron-impurity screening. Impurities with charge $\pm e$ are in the subcritical regime, so that screening effects due to the electron system and the media surrounding the graphene layer can be modelled by an effective charge or dielectric constant. Its long-range nature means that is does not scatter between the valleys and its position i.e. whether it is located at a lattice site or at the centre of a hexagon is largely irrelevant. In contrast the short-range impurity does scatter between the valleys and we consider such impurities situated at a lattice site. For an impurity located on an $A$ or $B$ sublattice site at the origin, the correction to the Hamiltonian is
\begin{equation}
\label{eq-vmat}
\mathrm{V}_A(\mathbf{r})=V_0\left( \begin{array}{cccc}
\delta\left(\mathbf{r}\right) & 0 & \delta\left(\mathbf{r}\right) & 0 \\
0 & 0 & 0 & 0 \\
\delta\left(\mathbf{r}\right) & 0 & \delta\left(\mathbf{r}\right) & 0\\
0 & 0 & 0 & 0
\end{array}\right),
\quad
\mathrm{V}_B(\mathbf{r})=V_0\left( \begin{array}{cccc}
0 & 0 & 0 & 0 \\
0 & \delta\left(\mathbf{r}\right) & 0 & -\delta\left(\mathbf{r}\right)\omega^\ast \\
0 & 0 & 0 & 0\\
0 &-\delta\left(\mathbf{r}\right)\omega &0 & \delta\left(\mathbf{r}\right)
\end{array}\right),
\end{equation}
respectively, where $\omega=e^{2\pi i/3}$ \cite{AndN98}.
The Coulomb potential is diagonal in both the sublattice and valley indices. In contrast, the structure of Eq. \eqref{eq-vmat} imposes complex selection rules, determining which transitions are connected via the short-range impurity. 

We consider excitations where an electron is promoted from one of the uppermost filled states $\mathcal{N}_2$ to an empty state in a higher lying LL $\mathcal{N}_1$ leaving behind a hole. We
introduce operators of collective excitations as
$Q^{\dag}_{\mathcal{ N}_1  \mathcal{ N}_2 M_z }  =
          \sum_{m_1 , m_2 = 0}^{\infty}
           A_{\mathcal{N}_1 \mathcal{N}_2 M_z }(m_1,m_2) \,
          c^{\dag}_{\mathcal{N}_1 m_1} d^{\dag}_{\mathcal{N}_2 m_2}$,
where the hole representation, $c_{\mathcal{N} m} \rightarrow d^{\dag}_{\mathcal{N} m}$
and $c^{\dag}_{\mathcal{N} m} \rightarrow d_{\mathcal{N} m}$ is used for all filled levels and the expansion coefficients satisfy
$A_{\mathcal{N}_1 \mathcal{N}_2 M_z }(m_1,m_2) \sim \delta_{M_z |n_1| - m_1 - |n_2| + m_2}$. The general secondary quantised form of the Hamiltonian is
\begin{align}
\label{ham} 
\hat{H} & = \sum_{\mathcal{N}, \mathcal{N'}} \sum_{m, m'}\left( \delta_\mathcal{NN'}\delta_{mm'}\tilde{\epsilon}_\mathcal{N} +{\mathcal{V}_i}_{\mathcal{N}m}^{\mathcal{N'}m'}\right) \left( c^{\dag}_{\mathcal{N'} m'}c_{\mathcal{N} m}- d^{\dag}_{\mathcal{N'} m'}d_{\mathcal{N} m}\right) \\ 
& + \sum_{\substack{\mathcal{N}_1, \mathcal{N}_2 \\ \mathcal{N}_1', \mathcal{N}_2'} }\sum_{\substack{m_1, m_2 \\ m_1', m_2'} }\mathcal{W}_{\mathcal{N}_1 m_1   \mathcal{N}_2 m_2}^{\mathcal{N}_1' m_1'  \mathcal{N}_2' m_2'}c^{\dag}_{\mathcal{N}_1' m_1'}d^{\dag}_{\mathcal{N}_2' m_2'}d_{\mathcal{N}_2 m_2}c_{\mathcal{N}_1 m_1}.
\end{align}
The first term gives the single particle energies ($\tilde{\epsilon}_\mathcal{N}$), which are lowered by an exchange self energy correction together with the impurity interaction. The impurity matrix elements are given by ${\mathcal{V}_i }_{\mathcal{N} m}^{\mathcal{N'} m'}=\delta_{s s'}\int \int d \mathbf{r} ^2 \Phi^\dag_{ N'}(\mathbf{r})V_i\Phi_{ N}(\mathbf{r})$, where $i \in \left\lbrace A, B, \mathrm{C}\right\rbrace$ denotes the impurity type. The second term gives the electron-hole ($e$-$h$) interactions governed by the Coulomb potential, $U\left( |\mathbf{r}_1-\mathbf{r}_2|\right) =\frac{e^2}{\varepsilon |\mathbf{r}_1-\mathbf{r}_2|}$, where $\varepsilon$ is an effective dielectric constant for $e$-$e$ interactions in graphene. It is made up of the $e$-$h$ attraction and the exchange repulsion,
$\mathcal{W}_{\mathcal{N}_1 m_1   \mathcal{N}_2 m_2}^{\mathcal{N}_1' m_1'  \mathcal{N}_2' m_2'}=
-\mathcal{U}_{\mathcal{N}_1 m_1   \mathcal{N}_2' m_2'}^{\mathcal{N}_1' m_1'  \mathcal{N}_2 m_2}
+\mathcal{U}_{\mathcal{N}_1 m_1   \mathcal{N}_2' m_2'}^{\mathcal{N}_2 m_2  \mathcal{N}_1 m_1}$
where the definition of matrix element $\mathcal{U}$ is specified elsewhere \cite{FisDR09b}. 

In order to solve the problem numerically, we need to decide which transitions contribute significantly to the optical response of the system and to reduce the number of terms in (\ref{ham}) to a finite number. In our calculations we assume all LLs with $n<0$ are filled, all LLs with $n>0$ are empty and that the sublevels of the zeroth LL become successively completely filled ($\nu=1,2,3,4$). Our focus is on determining the optically bright localised collective excitations. Only single particle transitions with no spin or pseudospin flips, no change of oscillator quantum number and $|n_1| - |n_2|= \pm 1$ are optically active in the two circular polarizations $\sigma^{\pm}$. This means that only transitions with  $M_z=\pm 1$ are active in $\sigma^{\pm}$. 
It can be shown that the only mixing of transitions with no spin or pseudospin flip by the $e$-$h$ Coulomb interaction is via the exchange interaction to other transitions with no spin or pseudospin flip. We only include transitions which are in resonance with each other, neglecting the weak correction of higher order terms, so that for the filling described, only the transitions from LL $n=0$ to LL $n=1$ ($0\to1$ transitions) and $-1 \to 0$ transitions are considered. For the Coulomb impurity there are four transitions that must be included for each filling factor; the relevant transitions for $\nu=1$ are shown within the dashed box in Fig.\ \ref{fig-diag}. The situation for the $\delta$-function impurity is somewhat more complicated, since it mixes different transitions, which may have different $M_z$ quantum numbers. The number of relevant transitions changes with the filling factor, light polarisation and the sublattice, which the impurity is located on. An example of the important transitions for the $\sigma^+$ light polarisation, $\nu=1$ and an impurity on the $A$ sublattice is shown in Fig.\ \ref{fig-diag}. We diagonalise the Hamiltonian numerically and obtain the energies and eigenvectors of the bound states. This requires cutting off $m$ at a finite value, which is justified as we seek excitations which are localised on the impurity. 
\section{Results}
\label{sec-res}
\begin{figure}
\begin{minipage}{72mm}
\includegraphics[width=\linewidth,height=35mm]{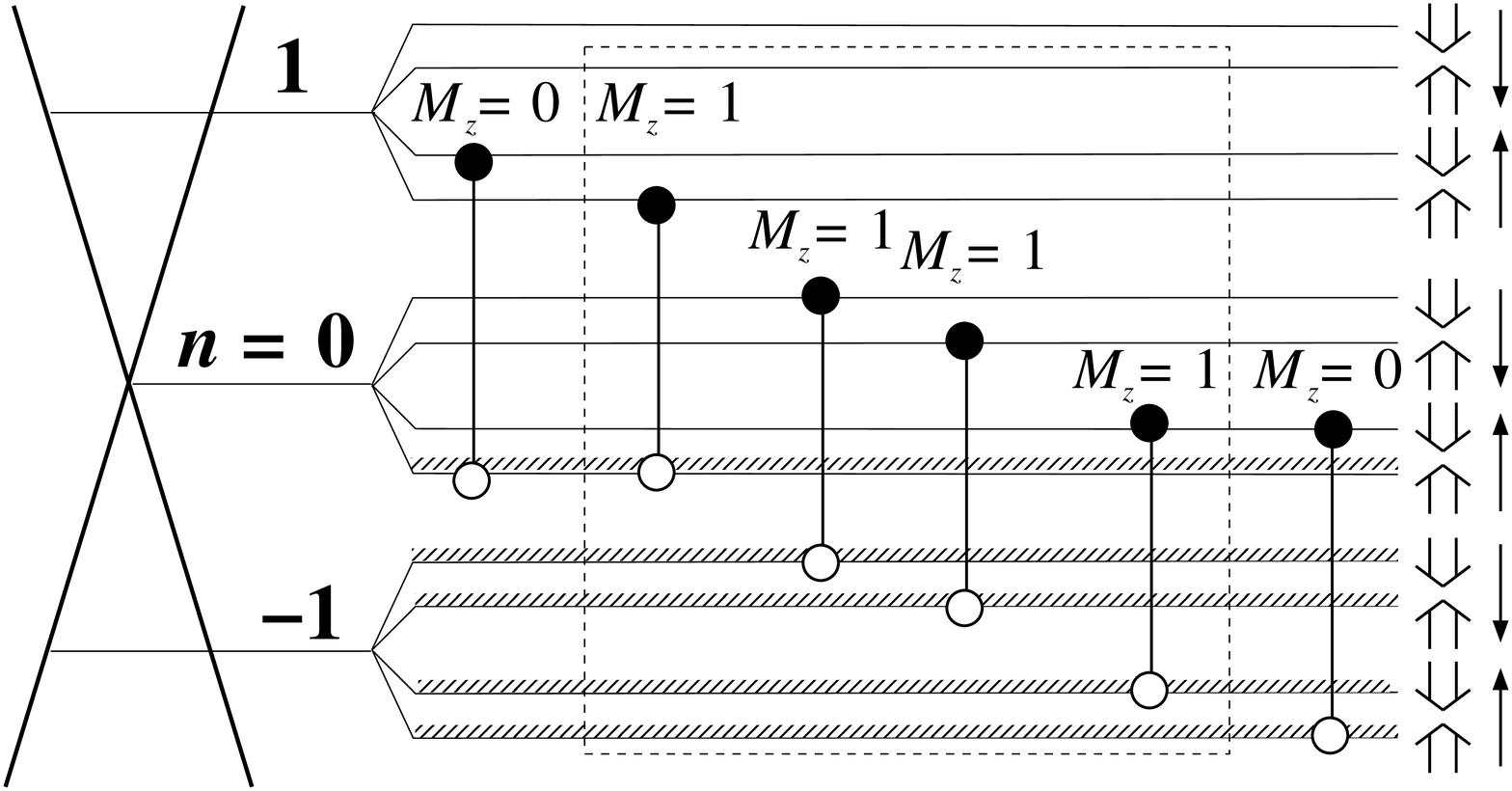}
\caption{Diagram indicating mixed excitations for a system with $\nu=1$ illuminated by $\sigma^+$ polarised light with a $\delta$-function impurity on the $A$ sublattice. The dashed box encloses the excitations which are mixed for the case of a Coulomb impurity.}
\label{fig-diag}
\end{minipage}
\hfil
\begin{minipage}{65mm}
\includegraphics[width=\linewidth,height=45mm]{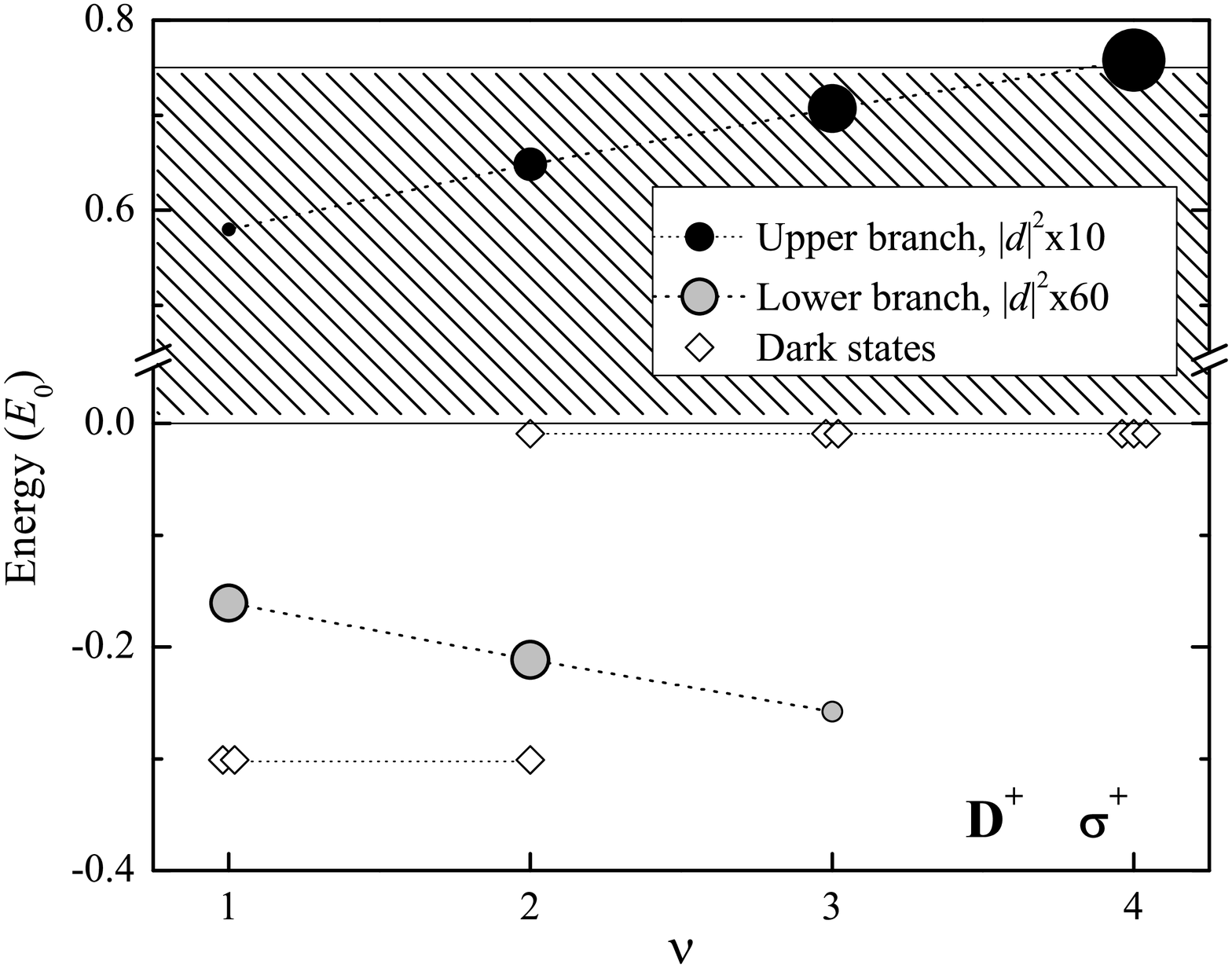}
\caption{Bound states plotted against the filling factor $\nu$ for a system illuminated by $\sigma^+$ polarised light with a charged donor Coulomb impurity, $D^+$. The dashed lines act as a guide to the eye. }
\label{fig-Coulomb}
\end{minipage}
\end{figure}

\begin{figure}
a)\includegraphics[width=68mm,height=45mm]{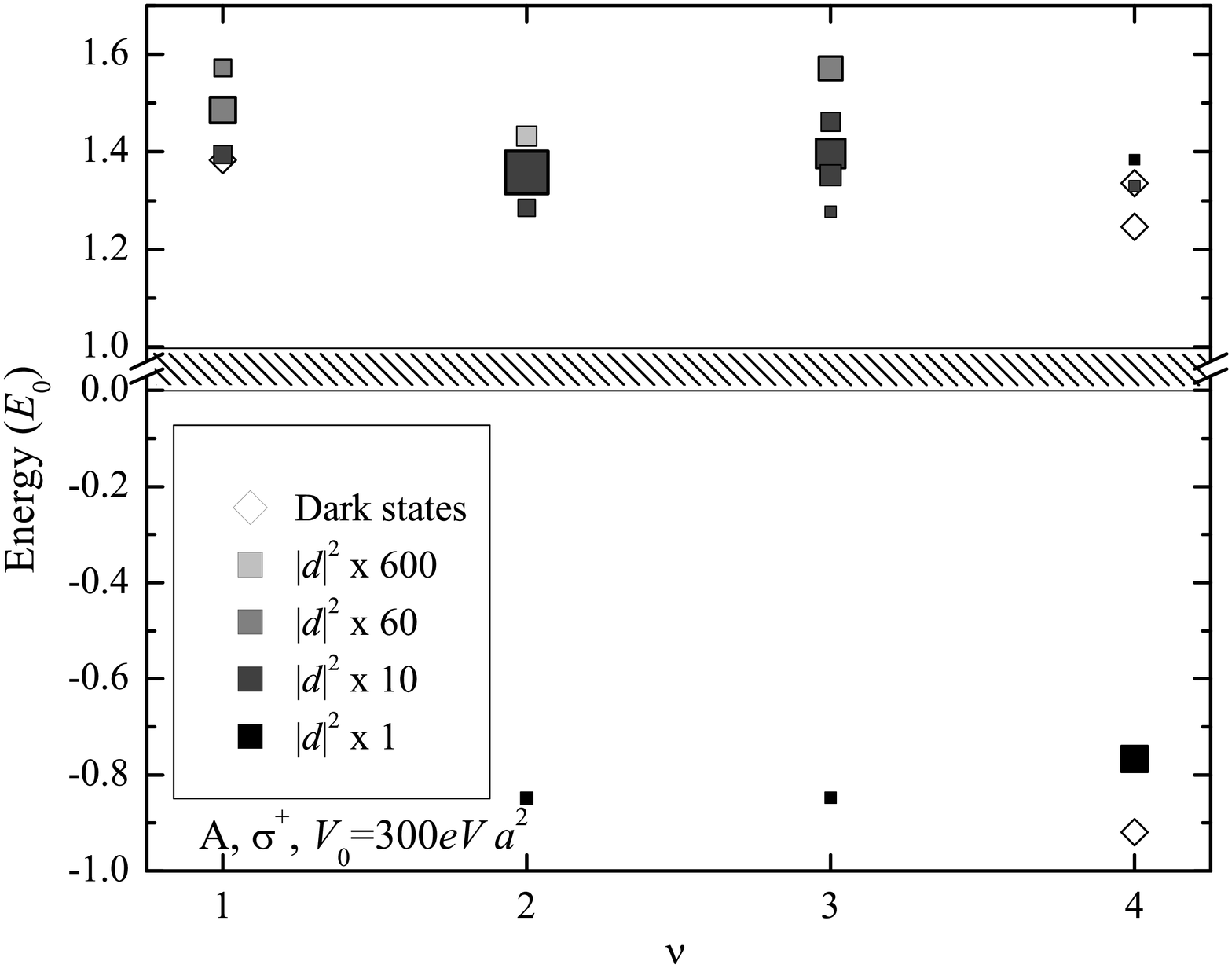}
\hfil
b)\includegraphics[width=68mm,height=45mm]{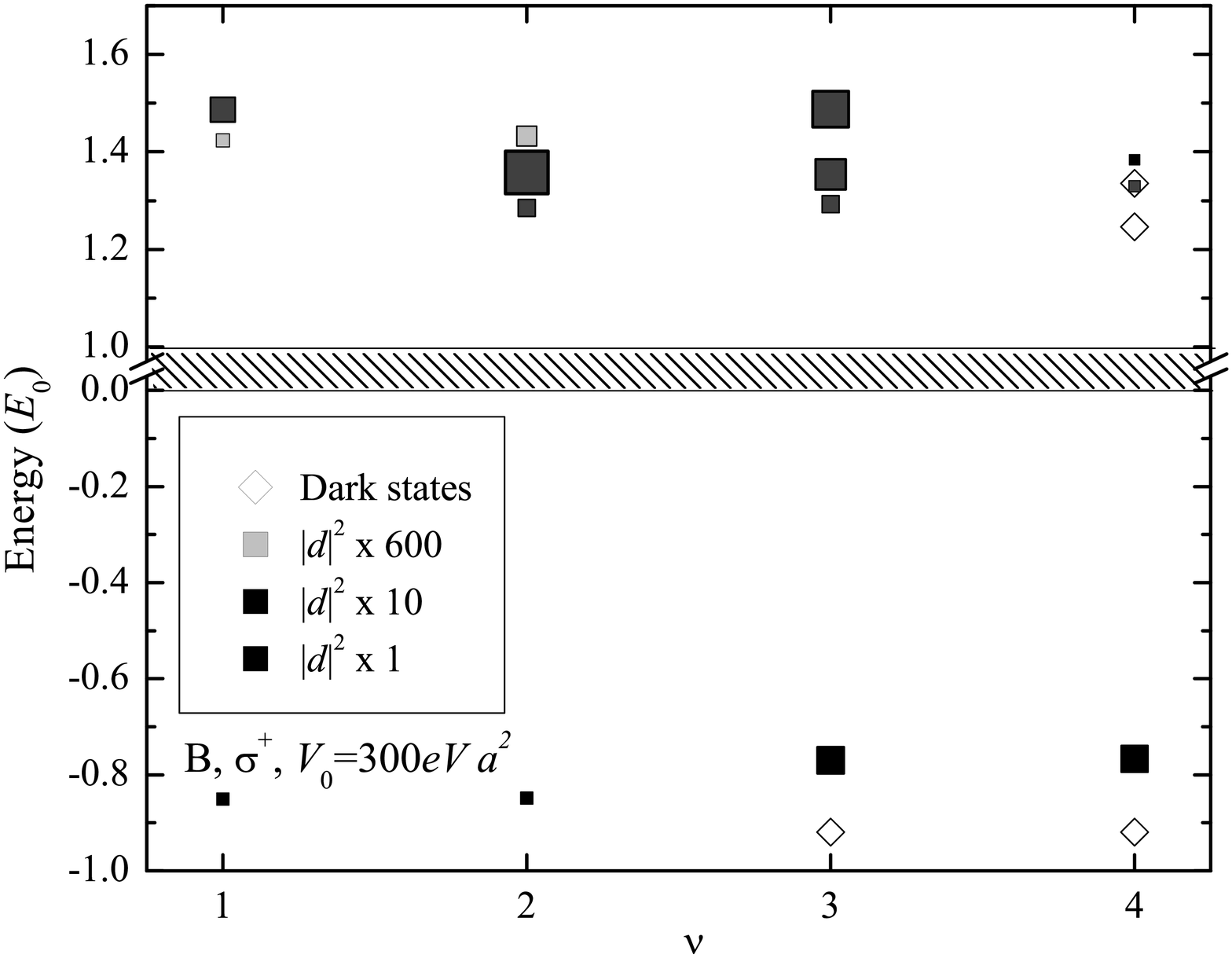}
\caption{Bound states plotted against the filling factor $\nu$ for a system illuminated by $\sigma^+$ polarised light with a $\delta$-function impurity on \textbf{a} the $A$ sublattice and \textbf{b} the $B$ sublattice. In real systems there will be a mixture of both.}
\label{fig-delta}
\end{figure}
In Figs.\ \ref{fig-Coulomb} and \ref{fig-delta}, we show the energies of the bound states (in units $E_0 = (\pi/2)^{1/2} e^2/\varepsilon \ell_B$, where $\ell_B$ is the magnetic length) plotted as a function of filling factor $\nu=1,2,3,4$ for the case of the Coulomb impurity and an impurity on the $A/B$ sublattices respectively. The Coulomb impurity is a donor $D^+$ with charge $+e$; the $\delta$-function impurity has strength $V_0=Wa^2$, where $a$ is the lattice constant and $W=300eV$. In both cases the system is illuminated by light in the $\sigma^+$ polarisation. In pristine graphene all the collective excitations are of an extended nature and possess a quasimomentum $K$. They are termed magnetoplasmons and form a continuum of width $\sim E_0$ contained here within the shaded area. The impurity results in some states becoming localised and splitting off from the magnetoplasmon continuum. For the case of the Coulomb impurity, we have also found quasibound states within the band, which have a high probability to exist on the impurity, but also long-range oscillating tails. Signatures of such states were also seen for the $\delta$-function impurity, but were not as pronounced. The symbol size in Figs.\ \ref{fig-Coulomb} and \ref{fig-delta} is proportional to the oscillator strength ($|d^2|$), with a magnification factor indicated in the legend; diamonds represent the dark states. Notice the differences between the graphs for the short-range impurity on the $A$ sublattice and that on the $B$ sublattice. The results are the same for even filling factors, but not for odd filling factors. This is because the valley splitting means that for odd filling factors the valleys are not equally filled making them inequivalent, which in turn introduces an inequivalence between the $A$ and $B$ sublattices. A significant difference between the impurity types is that there are fewer states localised on the Coulomb impurity and they evolve more smoothly as a function of the filling factor than those localised on a short-range scatterer. They form branches, which are indicated by the dashed lines in Fig.\ \ref{fig-Coulomb} with oscillator strengths which behave monotonically with increasing filling factor. The reason for this difference in behaviour is rooted in the greater complexity of the transitions involved for the $\delta$-function impurity and the fact that the number of relevant transitions and their $M_z$ values change as a function of filling factor $\nu$. In contrast, for the Coulomb impurity exactly four transitions are relevant for every filling factor each with $M_z=1$ and it is only the distribution between $0 \to 1$ and $-1 \to 0$ transitions that changes.

\section{Conclusions}
\label{sec-res}
We have studied the formation of optically-induced excitations bound on both a long-range (charged) and short-range (neutral) impurity in graphene in the presence of a strong magnetic field. The Coulomb interaction between different excitations mixes them, resulting in collective excitations of the whole system. In a real system, the short-range scatterer can be thought of as an impurity in the graphene layer at one of the lattice sites, whereas the Coulomb impurity is more likely to represent an impurity in the substrate very close to the graphene layer. For such states to be detected by magneto-optical spectroscopy, large clean samples are required so that the impurity density is low enough to be modelled by our single impurity model, whilst simultaneously the number of impurities is high enough to produce a detectable optical signature. However, with the current high focus of research in this area, we are confident that such samples will soon be available. It is important to note that the bound states due to a Coulomb impurity are more likely to be experimentally detectable, since the $\delta$-function impurity is less effective at producing bound states (they are only formed at high impurity strengths), as detailed in a separate work.
\providecommand{\WileyBibTextsc}{}
\let\textsc\WileyBibTextsc
\providecommand{\othercit}{}
\providecommand{\jr}[1]{#1}
\providecommand{\etal}{~et~al.}

\end{document}